\documentstyle[aps]{revtex}
\newcommand{\be}{\begin{equation}}
\newcommand{\ee}{\end{equation}}
\newcommand{\ba}{\begin{eqnarray}}
\newcommand{\ea}{\end{eqnarray}}
\newcommand{\baa}{\begin{eqnarray*}}
\newcommand{\eaa}{\end{eqnarray*}}

\newcommand{\dis}{\displaystyle}
\newcommand{\biq}{\mbox{\boldmath $q$}}
\newcommand{\bip}{\mbox{\boldmath $p$}}
\newcommand{\bif}{\mbox{\boldmath $f$}}

\begin{document}
\draft
{\pagestyle{empty}
\vskip 1.5cm

\title{Replica-exchange multicanonical algorithm and multicanonical
replica-exchange method for simulating systems with rough energy landscape}

\maketitle
\vskip 0.5cm

\centerline{Yuji Sugita\footnote{\ \ e-mail: sugita@ims.ac.jp}
and Yuko Okamoto\footnote{\ \ e-mail: okamotoy@ims.ac.jp}}
\centerline{{\it Department of 
Theoretical Studies, Institute for Molecular Science}}
\centerline{{\it Okazaki, Aichi 444-8585, Japan}}
\centerline{and}
\centerline{{\it Department of Functional Molecular Science,
The Graduate University for Advanced Studies}}
\centerline{{\it Okazaki, Aichi 444-8585, Japan}}

\vskip 2.0cm
\centerline{{\it Chem. Phys. Lett.}, in press.}

\begin{abstract}
 We propose two efficient algorithms for configurational sampling
 of systems with rough energy landscape. The first one is a new
 method for the determination of the multicanonical weight factor.
 In this method a short replica-exchange simulation is performed
 and the multicanonical weight factor is obtained by the multiple-histogram
 reweighting techniques.
 The second one is a further extension of the first in which
 a replica-exchange multicanonical simulation is performed
 with a small number of replicas.
 These new algorithms
 are particularly useful for studying the protein folding problem.
\end{abstract}
}

\section{Introduction}
With the great advancement of computational science over the
past decades, partly due to the rapid growth of computer
technologies, scientists in many research fields are now 
attacking realistic problems with large degrees of freedom
and complexity.  When dealing with systems with rough
energy landscape, simulations by conventional
molecular dynamics (MD) or Monte Carlo (MC) methods
are of little use, because at low temperatures
they tend to get trapped in
local-minimum energy states and sample only a very
limited region of the configurational space.
The development of new powerful simulation algorithms that
can alleviate this difficulty is
thus particularly important.

{\it Multicanonical algorithm} (MUCA) \cite{MUCA1}
may be one of the most popular methods that can sample
a wide phase space (for a review, see, e.g., Ref.~\cite{MUCArev}).  
A simulation in multicanonical ensemble
is based on a non-Boltzmann weight factor, which we refer to
as the multicanonical weight factor, and a free
random walk in potential energy space is realized. 
The random walk can overcome any energy barrier and 
allows the simulation to escape from local-minimum states.
Monitoring the potential energy throughout a MUCA
simulation, one can obtain not only the global-minimum energy
state but also canonical-ensemble averages of any physical
quantity as a function of temperature.  The latter is
accomplished by the single-histogram reweighting techniques
\cite{FS1}.
The multicanonical weight factor is usually determined by
iterations of short trial simulations (the details of this process
are described, for instance, in Refs.~\cite{MUCA3,HO94}).
This iterative process can be very difficult and
tedious for
complex systems, and there have been attempts to accelerate
the convergence of the iterative process 
\cite{MUCAIT2}--\cite{MUCAMD3}.
Despite these efforts, the determination of the multicanonical weight factor
still remains to be non-trivial \cite{MUCArev}.
  
Another powerful algorithm is 
the {\it replica-exchange method} (REM) 
\cite{RE1,RE2}. (A similar method was independently developed
earlier in Ref.~\cite{RE3}. 
REM is also referred to as {\it multiple Markov chain method} \cite{RE4}
and {\it parallel tempering} \cite{RE5}.) 
In this method, a number of
non-interacting copies of the original system (or replicas) 
at different temperatures are
simulated independently and
simultaneously by the conventional MD or MC methods. Every few steps,
pairs of replicas are exchanged with a specified transition
probability.
This exchange process realizes a random walk
in temperature space, which in turn induces
a random walk in the energy space so that a wide configurational
space can be sampled during the simulation.
The multiple-histogram reweighting techniques \cite{FS2,WHAM}
(an extension of which is also referred to as WHAM \cite{WHAM})
are used to calculate various canonical-ensemble averages
as a function of temperature.
In the replica-exchange method the weight factor is just the
product of Boltzmann factors, and so it is essentially known.
However, REM also has a
weak point:  It is difficult to study the
first-order phase transitions, because the replica-exchange rates 
passing the
first-order transition temperatures are greatly reduced.

In a previous work we have developed a molecular dynamics algorithm
for REM \cite{SO} (see, also, Refs.~\cite{H97,Yama}).  
We then developed a multidimensional REM \cite{SKO}
(see, also, Refs.~\cite{Huk2,YP}).
In this Letter we try to combine the
merits of multicanonical algorithm and replica-exchange method.
In the first method, which we refer to as
{\it replica-exchange multicanonical algorithm} (REMUCA),
a short replica-exchange simulation is performed and
the multicanonical weight factor is determined by
the multiple-histogram reweighting techniques \cite{FS2,WHAM}.
(We remark that the multiple-histogram reweighting techniques 
are also used during the regular 
iterations of multicanonical weight factor
determination in Refs.~\cite{MUCAIT2,MUCAMD3}).
We then present a further extension of the new method,
 which we refer to as {\it multicanonical replica-exchange method}
 (MUCAREM),
 where a replica-exchange multicanonical simulation is performed
 with a small number of replicas.
The effectiveness of these methods is tested with a short 
peptide, Met-enkephalin, in gas phase. 

\section{Methods}

\subsection{Replica-exchange multicanonical algorithm}
In the multicanonical ensemble \cite{MUCA1,MUCArev}
each state is weighted by the {\it multicanonical
weight factor}, $W_{mu}(E)$,
so that a flat potential energy
distribution, $P_{mu}(E)$, is obtained:
$P_{mu}(E) \propto n(E) W_{mu}(E) \equiv {\rm const}~,$
where $E$ is the potential energy and
$n(E)$ is the density of states.
We can then write
\begin{equation}
 W_{mu}(E) \equiv e^{-\beta_0 E_{mu}(E;T_0)}
= \frac{1}{n(E)}~, 
\label{eqn6}
\end{equation}
where we have chosen an arbitrary reference 
temperature, $T_0 = 1/k_B \beta_0$
($k_B$ is Boltzmann's constant), and
the ``{\it multicanonical potential energy}''
is defined by
\begin{equation}
 E_{mu}(E;T_0) = k_B T_0 \ln n(E)~.
\label{eqn7}
\end{equation}
Since the density of states of the system is usually unknown,
the multicanonical weight factor 
has to be numerically determined by iterations of short preliminary 
runs \cite{MUCA1,MUCArev}.

Once the weight factor is given, multicanonical Monte Carlo simulation
is performed, for instance, with the Metropolis criterion.
The molecular dynamics algorithm in multicanonical ensemble
also naturally follows from Eq.~(\ref{eqn6}), in which the
regular constant temperature molecular dynamics
(with $T=T_0$) is performed by solving the following modified
Newton equation: \cite{MUCAMD1,MUCAMD2,MUCAMD3} 
\begin{equation}
\dot{{\bip}}_i ~=~ - \frac{\partial E_{mu}(E;T_0)}{\partial {\biq}_i}
~=~ \frac{\partial E_{mu}(E;T_0)}{\partial E}~{\bif}_i~,
\label{eqn9a}
\end{equation}
where ${\biq}_i$ and ${\bip}_i$ are the coordinate vector and
momentum vector of the $i$-th atom, and
${\bif}_i$ is the regular force acting on the $i$-th atom.
In this Letter the formulations of the new methods in terms
of molecular dynamics algorithm are presented.
The details of the
Monte Carlo versions of the new algorithms have also been
worked out and will be given elsewhere \cite{MSO}.

We now briefly review the original replica-exchange method (REM)
\cite{RE1}-\cite{RE5} (see Ref.~\cite{SO} for details).
The system for REM consists of
$M$ {\it non-interacting} copies (or, replicas)
of the original system in the canonical ensemble
at $M$ different temperatures $T_m$ ($m=1, \cdots, M$).
Let $X = \left\{\cdots, x_m^{[i]}, \cdots \right\}$
stand for a state in this generalized ensemble.
Here, the superscript $i$ and the subscript $m$ in $x_m^{[i]}$
label the replica and the temperature, respectively.
The state $X$ is specified by the $M$ sets of
coordinates $q^{[i]}$ (and momenta $p^{[i]}$).

A simulation of the
{\it replica-exchange method} (REM)
is then realized by alternately performing the following two
steps \cite{RE1}-\cite{RE5}.
Step 1: Each replica in canonical ensemble of the fixed temperature
is simulated $simultaneously$ and $independently$
for a certain MC or MD steps.
Step 2: A pair of replicas,
say $i$ and $j$, which are
at neighboring temperatures $T_m$ and $T_n$, respectively,
are exchanged:
$X = \left\{\cdots, x_m^{[i]}, \cdots, x_n^{[j]}, \cdots \right\}
\longrightarrow \
X^{\prime} = \left\{\cdots, x_m^{[j]}, \cdots, x_n^{[i]},
\cdots \right\}$.
The transition probability of this replica exchange is given by
the Metropolis criterion:
\begin{equation}
w(X \rightarrow X^{\prime}) \equiv
w\left( x_m^{[i]} ~\left|~ x_n^{[j]} \right. \right)
= \left\{
\begin{array}{ll}
 1~, & {\rm for} \ \Delta \le 0~, \cr
 \exp \left( - \Delta \right)~, & {\rm for} \ \Delta > 0~,
\end{array}
\right.
\label{eqn24}
\end{equation}
where
\begin{equation}
\Delta = \left(\beta_m - \beta_n \right)
\left(E\left(q^{[j]}\right) - E\left(q^{[i]}\right)\right)~.
\label{eqn23b}
\end{equation}
Here, 
$E\left(q^{[i]}\right)$ and $E\left(q^{[j]}\right)$
are the potential energy of the $i$-th replica
and the $j$-th replica, respectively.
In the present work we employ molecular dynamics algorithm
for Step 1.
We remark that in molecular dynamics simulations we also
have to deal with the momenta.
It was shown that by
rescaling the momenta uniformly 
by the square root of the ratio of
$T_m$ and $T_n$,
the kinetic energy terms in the Boltzmann factors 
are cancelled out and that
the same criterion, Eqs.~(\ref{eqn24}) and (\ref{eqn23b}),
which was originally
derived for Monte Carlo algorithm \cite{RE1}-\cite{RE5}
is recovered \cite{SO}.

The major advantage of REM over other generalized-ensemble
methods such as multicanonical algorithm \cite{MUCA1}
lies in the fact that the weight factor
is essentially {\it a priori} known,
while in the latter algorithm the determination of the
weight factor can be very tedious and time-consuming.
A random walk in ``temperature space'' is
realized for each replica, which in turn induces a random
walk in potential energy space.  This alleviates the problem
of getting trapped in states of energy local minima.

In the {\it replica-exchange multicanonical algorithm} (REMUCA) 
we first perform a short REM simulation (with $M$ replicas)
to determine the 
multicanonical weight factor and then perform with this weight
factor a regular multicanonical simulation with high statistics.
The first step is accomplished by the multiple-histogram reweighting
techniques \cite{FS2,WHAM}.
Let $N_m(E)$ and $n_m$ be respectively
the potential-energy histogram and the total number of
samples obtained at temperature $T_m=1/k_B \beta_m$ of the REM run.
The density of states, $n(E)$,
is then given by \cite{FS2,WHAM}
\begin{equation}
n(E) = \frac{\dis{\sum_{m=1}^M g_m^{-1}~N_m(E)}}
{\dis{\sum_{m=1}^M n_{m}~g_m^{-1}~e^{f_m-\beta_m E}}}~,
\label{eqn27}
\end{equation}
and
\begin{equation}
e^{-f_m} = \sum_{E} n(E) e^{-\beta_m E}~.
\label{eqn28}
\end{equation}
Here, $g_m = 1 + 2 \tau_m$,
and $\tau_m$ is the integrated
autocorrelation time at temperature $T_m$.
Note that the density of states $n(E)$ and the ``dimensionless''
Helmholtz free energy $f_m$ in Eqs.~(\ref{eqn27}) and
(\ref{eqn28}) are solved self-consistently
by iteration \cite{FS2,WHAM}.

Once the estimate of the density of states is obtained, the 
multicanonical weight factor can be directly determined by using
Eq.~(\ref{eqn6}) (see also Eq.~(\ref{eqn7})).
Actually, the multicanonical potential energy, $E_{mu}(E;T_0)$, 
thus determined is only reliable in the following range:
\begin{equation}
E_1 \le E \le E_M~,
\label{eqn29}
\end{equation}
where $T_1$ and $T_M$ are respectively the lowest and the highest 
temperatures used in the REM run and
\begin{equation}
\left\{
\begin{array}{rl}
E_1 &=~ <E>_{T_1}~, \\
E_M &=~ <E>_{T_M}~.
\end{array}
\right.
\label{eqn29b}
\end{equation}
Here, $<E>_T$ stands for the canonical expectation value of 
$E$ at temperature $T$.
Outside this range we extrapolate
the multicanonical potential energy linearly as follows:
\begin{equation}
 {\cal E}_{mu}^{\{0\}}(E) \equiv \left\{
   \begin{array}{@{\,}ll}
   \left. \dis{\frac{\partial E_{mu}(E;T_0)}{\partial E}}
        \right|_{E=E_1} (E - E_1)
             + E_{mu}(E_1;T_0)~, &
         \mbox{for $E < E_1$~,} \\
         E_{mu}(E;T_0)~, &
         \mbox{for $E_1 \le E \le E_M$~,} \\
   \left. \dis{\frac{\partial E_{mu}(E;T_0)}{\partial E}}
        \right|_{E=E_M} (E - E_M)
             + E_{mu}(E_M;T_0)~, &
         \mbox{for $E > E_M$~.}
   \end{array}
   \right. 
\label{eqn31}
\end{equation}
In the present work the multicanonical potential
energy function, $E_{mu}(E;T_0)$, and its derivative,
$\frac{\partial E_{mu}(E;T_0)}{\partial E}$, were obtained by 
fitting the logarithm of the density of states, $\ln n(E)$, (which was
determined by the multiple-histogram reweighting
techniques) by the cubic
spline functions in the energy range of Eq.~(\ref{eqn29}).
A long multicanonical MD run is then performed by solving
the Newton equations in Eq.~(\ref{eqn9a}) 
into which we substitute ${\cal E}_{mu}^{\{0\}}(E)$ of
Eq.~(\ref{eqn31}).

Let $N_{mu}(E)$ be the histogram of the
potential energy distribution, $P_{mu}(E)$, that is 
obtained from this multicanonical production run.
The expectation value of a physical quantity $A$
at any temperature $T$ is then calculated from
\begin{equation}
<A>_{T} \ = \frac{\dis{\sum_{E}A(E)~n(E)~e^{-\beta E}}}
{\dis{\sum_{E} n(E)~e^{-\beta E}}}~,
\label{eqn18}
\end{equation}
where the (unnormalized) density of states is obtained by 
the single-histogram
reweighting techniques \cite{FS1}:
\begin{equation}
 n(E) = \frac{N_{mu}(E)}{W_{mu}(E)}~.
\label{eqn17}
\end{equation}

We remark that
our multicanonical MD simulation here
actually results in a canonical simulation at
$T=T_1$ for $E < E_1$, a multicanonical simulation for
$E_1 \le E \le E_M$, and a canonical simulation at
$T=T_M$ for $E > E_M$ (a detailed discussion on this
point will be given elsewhere \cite{SO4}).
Note also that the above arguments are independent of
the value of $T_0$, and we
will get the same results, regardless of its value.

Finally, although we did not find any difficulty in the case of protein systems,
a single REM run in general may not be able to 
give an accurate estimate of the
density of states (like in the case of
a first-order phase transition).  In such a
case we can still greatly simplify the process of the
multicanonical weight factor determination by
combining the present method with the
previous iterative methods \cite{MUCAIT2}-\cite{MUCAMD3}.

\subsection{Multicanonical replica-exchange method}
In the previous subsection we presented a new generalized-ensemble
algorithm, REMUCA, 
that uses the replica-exchange method
for the determination of the multicanonical weight factor.  
Here, we present a further modification of REMUCA and refer the
new method as {\it multicanonical replica-exchange method}
(MUCAREM).  In MUCAREM the final production run is not a regular
multicanonical simulation (as in REMUCA)
but a replica-exchange simulation
with a few replicas, say ${\cal M}$ replicas,
in the multicanonical ensemble. 
We expect that this enhances the sampling further because
a replica-exchange process can be considered to be a
global update,
while multicanonical simulations are usually based on local
updates.
We remark that replica-exchange simulations based on another
generalized ensemble are introduced in Ref.~\cite{H97}.

We now give the details of MUCAREM.
As in REMUCA, we first perform a short REM simulation with $M$ replicas
with $M$ different temperatures (we order them as 
$T_1 < T_2 < \cdots < T_M$)
and obtain the best estimate of the density of states $n(E)$
in the whole energy range of interest (see Eq.~(\ref{eqn29})).
We then choose a number ${\cal M}$ (${\cal M} \ll M$) and
assign ${\cal M}$ pairs of temperatures 
($T_L^{\{m\}}, T_H^{\{m\}}$) ($m = 1, \cdots, {\cal M}$).
Here, we assume that
$T_L^{\{m\}} < T_H^{\{m\}}$ and arrange the temperatures 
so that the neighboring regions covered by the pairs
have sufficient overlaps.  We also set  
$T_L^{\{1\}} = T_1$ and $T_H^{\{{\cal M}\}} = T_M$.
We choose ${\cal M}$ (arbitrary) temperatures
$T_m$ ($m = 1, \cdots, {\cal M}$).
We then define the following quantities:
\begin{equation}
\left\{
\begin{array}{rl}
E_L^{\{m\}} &=~ <E>_{T_L^{\{m\}}}~, \\
E_H^{\{m\}} &=~ <E>_{T_H^{\{m\}}}~,~(m = 1, \cdots, {\cal M})
\end{array}
\right.
\label{eqn32}
\end{equation}
and assign the following multicanonical potential energies
(see Eq.~(\ref{eqn31})):
\begin{equation}
 {\cal E}_{mu}^{\{m\}}(E) = \left\{
   \begin{array}{@{\,}ll}
   \left. \dis{\frac{\partial E_{mu}(E;T_m)}{\partial E}}
        \right|_{E=E_L^{\{m\}}} (E - E_L^{\{m\}})
             + E_{mu}(E_L^{\{m\}};T_m)~, &
         \mbox{for $E < E_L^{\{m\}}$,} \\
         E_{mu}(E;T_m)~, &
         \mbox{for $E_L^{\{m\}} \le E \le E_H^{\{m\}}$,} \\
   \left. \dis{\frac{\partial E_{mu}(E;T_m)}{\partial E}}
        \right|_{E=E_H^{\{m\}}} (E - E_H^{\{m\}})
             + E_{mu}(E_H^{\{m\}};T_m)~, &
         \mbox{for $E > E_H^{\{m\}}$,}
   \end{array}
   \right. 
\label{eqn33}
\end{equation}
where $E_{mu}(E;T)$ is the multicanonical potential energy 
that was determined for the whole energy range
of Eq.~(\ref{eqn29}).
The Newton equation in Eq.~(\ref{eqn9a}) again implies that 
our choice of ${\cal E}_{mu}^{\{m\}}(E)$ 
in Eq.~(\ref{eqn33}) results in a canonical simulation at
$T=T_L^{\{m\}}$ for $E < E_L^{\{m\}}$, a multicanonical simulation for
$E_L^{\{m\}} \le E \le E_H^{\{m\}}$, and a canonical simulation at
$T=T_H^{\{m\}}$ for $E > E_H^{\{m\}}$ \cite{SO4}.

The production run of MUCAREM is a replica-exchange simulation
with ${\cal M}$ replicas with ${\cal M}$ different
temperatures $T_m$ and multicanonical potential
energies ${\cal E}_{mu}^{\{m\}}(E)$.
By following the same derivation that led to the original
REM, one can show that the
transition probability of replica exchange, 
$w\left( x_m^{[i]} ~\left|~ x_n^{[j]} \right. \right)$,
is given by Eq.~(\ref{eqn24}), 
where we now have
\begin{equation}
\Delta = \beta_m
\left({\cal E}_{mu}^{\{m\}}\left(E\left(q^{[j]}\right)\right) -
{\cal E}_{mu}^{\{m\}}\left(E\left(q^{[i]}\right)\right)\right)
- \beta_n
\left({\cal E}_{mu}^{\{n\}}\left(E\left(q^{[j]}\right)\right) -
{\cal E}_{mu}^{\{n\}}\left(E\left(q^{[i]}\right)\right)\right)~.
\label{Eqn21}
\end{equation}
Note that we need to newly evaluate the multicanonical
potential energy, ${\cal E}_{mu}^{\{m\}}(E(q^{[j]}))$ and
${\cal E}_{mu}^{\{n\}}(E(q^{[i]}))$, because 
${\cal E}_{mu}^{\{m\}}(E)$ and
${\cal E}_{mu}^{\{n\}}(E)$ are, in
general, different functions for $m \ne n$. 
We remark that the same additional evaluation of the
potential energy is necessary for the multidimensional replica-exchange
method \cite{SKO}.

Let $N_m(E)$ and $n_m$ be respectively
the potential-energy histogram and the total number of
samples obtained at $T_m$ ($m = 1, \cdots, {\cal M}$).
The expectation value of a physical quantity $A$ 
at any temperature $T=1/k_B \beta$, $<A>_{T}$,
is then given by Eq.~(\ref{eqn18}),
where the density of states is now obtained from the
multiple-histogram reweighting formulae \cite{FS2,WHAM}
of Eqs.~(\ref{eqn27}) and (\ref{eqn28}) 
with the replacements:
$M \rightarrow {\cal M}$ and 
$\beta_m E \rightarrow \beta_m {\cal E}_{mu}^{\{m\}}(E)$.

\section{Results and discussion}
We tested the effectiveness of the new algorithms for the system of
a penta-peptide, Met-enkephalin, in gas phase. This peptide has the
amino-acid sequence, Tyr-Gly-Gly-Phe-Met. The N and C termini of the
peptide were blocked with acetyl and N-methyl groups, respectively. The
force-field parameters were taken from the all-atom version of AMBER
developed by Cornell {\it et al.} \cite{AM94}, and the dielectric 
constant was set equal to 1.0.

We have performed MD simulations in the Cartesian coordinates 
based on the replica-exchange multicanonical algorithm (REMUCA)
and multicanonical replica-exchange method (MUCAREM). 
As a thermostat for both MD simulations in the replica-exchange 
method (REM) and multicanonical algorithm (MUCA), we have 
adopted the constraint method \cite{HLM,EM} in which the
total kinetic energy is constant, following Ref.~\cite{MUCAMD2}
where the constraint method is recommended for MUCA simulations. 
The computer code developed in
Refs. \cite{SK,KHG}, which is based on PRESTO
\cite{PRESTO}, was used. The unit time step was set to 0.5 fs.
The simulations (of all replicas)
were started from an extended conformation.

In Table I we summarize the parameters of the simulations that
were performed in the present work. 
As discussed in the previous section, REMUCA consists of 
two simulations: a short REM simulation (from which the 
density of states of the system, or the multicanonical weight factor,
is determined) and a subsequent
production run of MUCA simulation.
The former simulation is referred to as REM1 and the latter
as MUCA1 in Table I.
A production run of MUCAREM simulation is referred to as 
MUCAREM1 in Table I,
and it uses the same density of states that was obtained
from REM1.
Finally, a production run of the original REM simulation
was also performed for comparison and it is referred to as
REM2 in Table I.
The total simulation time for the three production runs
(REM2, MUCA1, and MUCAREM1) was all set equal (i.e., 5 ns).
Before taking the data, we made regular canonical MD simulations of 
100 ps (for each replica) for thermalization.  For replica-exchange
simulations an additional REM simulation of 100 ps was made for
further thermalization.
In REM1 and REM2 there exist
10 replicas with 10 different temperatures, ranging from
200 K to 1000 K as listed in Table I (i.e., $T_1 = 200$ K
and $T_M = T_{10} = 1000$ K).  For the present purpose of the
work, we believe that this range of temperature is sufficiently
wide.  The temperatures are distributed exponentially between
$T_1$ and $T_M$, following the optimal
distribution found in Ref \cite{SO}. 
In REM1 and REM2 a replica exchange was tried every 20 time
steps (or 10 fs) as in our previous work \cite{SO}. 

In REM1 each replica was simulated
by a constant temperature MD of $2 \times 10^5$ steps 
(or 100 ps). 
We first check whether the replica-exchange simulation
of REM1 indeed performed properly.
For an optimal performance of REM the acceptance ratios 
of replica exchange should be
sufficiently uniform and large (say, $> 10$ \%).
In Table II we list these quantities.
It is clear that both points are met (the values vary only
between 13 \% and 16 \%). 
Moreover, in Fig.~1 the canonical 
probability distributions obtained
at the chosen 10 temperatures from REM1
are shown.  We see that there
are enough overlaps between all pairs of neighboring
distributions, indicating
that there will be sufficient numbers of replica exchanges
between pairs of replicas (see Table II).
We did observe random walks in 
potential energy space between low energies and high energies.

After REM1, we obtained the density of states, $n(E)$, by the
multiple-histogram method \cite{FS2,WHAM} (namely, by
solving Eqs.~(\ref{eqn27}) and (\ref{eqn28}) self-consistently). 
For biomolecular systems the integrated autocorrelation times, 
$\tau_m$, in the reweighting formulae
can safely be set to be a constant \cite{WHAM}, and we
do so throughout the analyses in this section.
The density of states will give the average values of the
potential energy from Eq.~(\ref{eqn18}), and we found
\begin{equation}
\left\{
\begin{array}{rl}
E_1 &=~ <E>_{T_1} = -30 ~{\rm kcal/mol}~, \\
E_M &=~ <E>_{T_M} = 195 ~{\rm kcal/mol}~.
\end{array}
\right.
\label{eqn50}
\end{equation}
Then our estimate of the density of states is reliable
in the range $E_1 \le E \le E_M$.
The multicanonical potential energy ${\cal E}_{mu}^{\{0\}}(E)$
was thus determined for the three energy regions
($E < E_1$, $E_1 \le E \le E_M$, and $E > E_M$) from
Eq.~(\ref{eqn31}).
Namely, the multicanonical potential energy, $E_{mu}(E;T_0)$,
in Eq.~(\ref{eqn7}) and its derivative, 
$\frac{\partial E_{mu}(E;T_0)}{\partial E}$,
were determined by fitting $\ln n(E)$ by cubic
spline functions in the energy region of 
($-30 \le E \le 195$ kcal/mol). 
Here, we have set the arbitrary reference temperature to be $T_0 = 1000$ K.
Outside this energy region, $E_{mu}(E;T_0)$ was linearly
extrapolated as in Eq.~(\ref{eqn31}).

After determining the multicanonical weight factor,
we carried out a multicanonical MD simulation of $1 \times
10^7$ steps (or 5 ns) for data collection (MUCA1 in Table I).
In Fig. 2 the probability distribution obtained by MUCA1 is plotted. 
It can be seen that a good
flat distribution is obtained in the energy region
$E_1 \le E \le E_M$. 
In Fig. 2 the canonical probability distributions that
were obtained by the reweighting techniques at 
$T = T_1 = 200$ K and $T = T_M = 1000$ K are also 
shown (these results are essentially identical to
one another among
MUCA1, MUCAREM1, and REM2, as discussed below).
Comparing these curves with those of MUCA1 in
the energy regions $E < E_1$
and $E > E_M$ in Fig. 2, we confirm our claim in the
previous section that
MUCA1 gives canonical distributions at $T=T_1$ for
$E < E_1$ and at $T=T_M$ for $E > E_M$, whereas
it gives a multicanonical distribution for 
$E_1 \le E \le E_M$. 

In the previous
works of multicanonical simulations of Met-enkephalin in gas phase
(see, for instance, Refs.~\cite{HO,MUCAMD2,MHO}), at least several
iterations of trial simulations were required for the multicanonical
weight determination.
We emphasize that in the present case of REMUCA (REM1), only
one simulation was necessary to determine the optimal multicanonical
weight factor that can cover the energy region corresponding to
temperatures between 200 K and 1000 K.

From the density of states obtained by
REMUCA (i.e., REM1), 
we prepared the multicanonical weight factors (or the multicanonical
potential energies) for the MUCAREM simulation (see
Eq.~(\ref{eqn33})).  
The parameters
of MUCAREM1, such as energy bounds
$E_L^{\{m\}}$ and $E_H^{\{m\}}$ ($m=1, \cdots, {\cal M}$) are
listed in Table III.  The choices of 
$T_L^{\{m\}}$ and $T_H^{\{m\}}$ are,
in general, arbitrary, but significant overlaps between the
probability distributions of adjacent replicas are necessary.
The replica-exchange process in MUCAREM1 was tried every 200
time steps (or 100 fs). It is less frequent than in REM1 (or REM2). 
This is because we
wanted to ensure a sufficient time for system relaxation. 

In Fig.~3 the probability distributions of potential energy
obtained by MUCAREM1 are shown.  
As expected, we observe that the
probability distributions corresponding to the temperature
$T_m$ are essentially flat for the energy region
$E_L^{\{m\}} \le E \le E_H^{\{m\}}$, are of the canonical simulation at
$T=T_L^{\{m\}}$ for $E < E_L^{\{m\}}$, and are of the
canonical simulation at $T=T_H^{\{m\}}$ for $E > E_H^{\{m\}}$
($m=1, \cdots, {\cal M}$).
As a result, each distribution in MUCAREM is much broader than those
in the conventional REM (see Fig. 1) and a much smaller number of 
replicas are required in MUCAREM than in REM 
(${\cal M}=4$ in MUCAREM versus $M=10$ in REM).

The acceptance probabilities of replica exchange in MUCAREM1
are listed in Table IV. 
Relatively high probabilities of replica exchange in 
MUCAREM1 should be noted. In MUCAREM1, the exchange probability is given
by Eqs.~(\ref{eqn24}) and (\ref{Eqn21}). 
Suppose that a pair of replicas $i$ (at temperature $T_m$)
and $j$ (at temperature $T_n$)
are respectively in the energy ranges between 
$E_L^{\{m\}} \le E \le E_H^{\{m\}}$ and between
$E_L^{\{n\}} \le E \le E_H^{\{n\}}$.
Then Eq.~(\ref{Eqn21}) together with Eqs.~(\ref{eqn7}) and
(\ref{eqn33}) gives
\begin{eqnarray}
\Delta &=& \beta_m \left(
E_{mu} \left(E\left(q^{[j]}\right);T_m\right)
-E_{mu} \left(E\left(q^{[i]}\right);T_m\right)\right)
   - \beta_n \left(
E_{mu} \left(E\left(q^{[j]}\right);T_n\right)
-E_{mu} \left(E\left(q^{[i]}\right);T_n\right)\right) \nonumber \\
       &=& \ln n\left(E\left(q^{[j]}\right)\right) - 
                 \ln n\left(E\left(q^{[i]}\right)\right) -
           \ln n\left(E\left(q^{[j]}\right)\right) + 
                 \ln n\left(E\left(q^{[i]}\right)\right) \nonumber \\
       &=& 0~.
\end{eqnarray}
Thus, in this case the replica exchange is {\it always} accepted. 
This is the main reason for the high acceptance ratio in MUCAREM simulations.

In Fig. 4 the time series of potential energy for the
first 500 ps of REM2 (a), MUCA1 (b), and MUCAREM1 (c) are plotted. 
Note that in MUCA1 (Fig. 4(b))
some extra MD steps are required to escape from the 
lowest-energy region (from about 300 ps to 450 ps) and the
transition from the highest to the lowest energy occurs only once during 
the first 500 ps in MUCA1. 
This is presumably
due to the nature of local updates.  
In REM2 and MUCAREM1, such transitions 
occur twice or three times. This is because global updates by
the replica-exchange process make the random walk more efficient
(more quantitative comparisons of REM, MUCA, and MUCAREM
will be given elsewhere \cite{SO4}).

To check the validity of the canonical-ensemble expectation values
calculated by the new algorithms, we
compare the average potential energy as a function of temperature
in Fig. 5.
In REM2 and MUCAREM1 we used the multiple-histogram 
techniques \cite{FS2,WHAM}
(see Eqs.~(\ref{eqn27}) and (\ref{eqn28}) for REM2 and
the corresponding variants for MUCAREM1), whereas the
single-histogram method \cite{FS1} (see Eq.~(\ref{eqn17}))
was used in MUCA1. We can see a perfect coincidence of
these quantities among REM2, MUCA1, and MUCAREM1 in Fig. 5. 

\section{Conclusions}
In this Letter we have proposed two new algorithms for configurational
sampling of systems with rough energy landscape.
In the first method, which we refer to as replica-exchange multicanonical
algorithm (REMUCA), the multicanonical weight factor is determined from 
the results of a short replica-exchange simulation, and then a regular
multicanonical production run is made with this weight.
In the second method, which we refer to as multicanonical replica-exchange
method (MUCAREM), the multicanonical weight factor is determined as in
REMUCA, and then a replica-exchange multicanonical production run is
made.  The number of required replicas in MUCAREM is much smaller than
in the original replica-exchange method.
The new methods were tested with 
the system of a small peptide in gas phase. 
This is one of the smallest models of protein systems.
The multicanonical weight factor was indeed obtained by a single, short
replica-exchange simulation.
It may occur, however, that this first step may not be as easy in
more complicated systems such as a protein with explicit
solvent molecules. In such a case one can still simplify the multicanonical
weight factor determination by iteratively combining the present
algorithms (and also the previous methods).

\noindent
{\bf Acknowledgements}

The authors are grateful to Dr. A. Kitao of Kyoto University
for 
useful discussions on fitting procedures in the determination of
the multicanonical weight factor.
Our simulations were performed on the Hitachi and other computers at the
IMS Computer Center.  Part of our simulations were also performed on the
supercomputers at the National Institute of Genetics. 
This work is supported, in part, by a grant from the Research
for the Future Program of the Japan Society for the Promotion of
Science (JSPS-RFTF98P01101).


%
%
%
%
\begin{table}
 \caption{Summary of parameters in REM, REMUCA, and MUCAREM simulations.}
 \begin{center}
 \begin{tabular}{cccc}
   Run     & No. of replicas, $M$ & Temperature, $T_m$ [K] 
   ($m = 1, \cdots, M$) & MD steps\\
   \hline
   REM1    & 10  & 200, 239, 286, 342, 409, & $2 \times 10^5$ \\
           &     & 489, 585, 700, 836, 1000 & \\
   REM2    & 10  & 200, 239, 286, 342, 409, & $1 \times 10^6$ \\
           &     & 489, 585, 700, 836, 1000 & \\
   MUCA1   & 1   & 1000 & $1 \times 10^7$ \\ 
   MUCAREM1 & 4    & 375, 525, 725, 1000     & $2.5 \times 10^6$ \\
  \end{tabular}
 \end{center}
 \label{Tab1}
\end{table}

\begin{table}
 \caption{Acceptance ratios of replica exchange in REM1.}
 \begin{center}
 \begin{tabular}{cc}
   Pair of temperatures [K] & Acceptance ratio \\
   \hline
    200~~  $\longleftrightarrow$   ~~239 & 0.15 \\
    239~~  $\longleftrightarrow$   ~~286 & 0.16 \\
    286~~  $\longleftrightarrow$   ~~342 & 0.15 \\
    342~~  $\longleftrightarrow$   ~~409 & 0.14 \\
    409~~  $\longleftrightarrow$   ~~489 & 0.13 \\
    489~~  $\longleftrightarrow$   ~~585 & 0.13 \\
    585~~  $\longleftrightarrow$   ~~700 & 0.13 \\
    700~~  $\longleftrightarrow$   ~~836 & 0.14 \\
    836~~  $\longleftrightarrow$   ~~1000 & 0.15 \\
  \end{tabular}
 \end{center}
 \label{Tab2}
\end{table}

\begin{table}
 \caption{Summary of parameters in MUCAREM1.}
 \begin{center}
 \begin{tabular}{ccccc}
    $m$ & $T_L^{\{m\}}$ [K] & $T_H^{\{m\}}$ [K]
& $E_L^{\{m\}}$ [kcal/mol] & $E_H^{\{m\}}$ [kcal/mol] \\
   \hline
    1 & 200 & 375  & $-30$ & 20 \\
    2 & 300 & 525  & $-5$  & 65 \\
    3 & 375 & 725  & 20  & 120 \\
    4 & 525 & 1000 & 65  & 195 \\
  \end{tabular}
 \end{center}
 \label{Tab3}
\end{table}

\begin{table}
 \caption{Acceptance ratios of replica exchange MUCAREM1.}
 \begin{center}
 \begin{tabular}{cc}
    Pair of temperatures [K] & Acceptance ratio \\
   \hline
    375~~  $\longleftrightarrow$  ~~525  & 0.25 \\
    525~~  $\longleftrightarrow$  ~~725  & 0.35 \\
    725~~  $\longleftrightarrow$  ~~1000 & 0.32 \\
  \end{tabular}
 \end{center}
 \label{Tab4}
\end{table}
\newpage

\centerline{\bf Figure Captions}

\begin{itemize}
 \item Fig.~1. Probability distribution of potential energy
               obtained from REM1 (see Table I for the parameters
               of the simulation).
 \item Fig.~2. Probability distribution of potential energy
        obtained from MUCA1
        (see Table I).  The dotted curves are the probability
       distributions of the reweighted
       canonical ensemble at $T = 200$ K (left) and 1000 K (right).

 \item Fig.~3. Probability distributions of potential energy
       obtained from MUCAREM1 (see Tables I and III).

 \item Fig.~4. Time series of potential energy for one of the
       replicas in (a) REM2, (b) MUCA1, and (c) MUCAREM1 
               (see Tables I and III for the parameters
               of the simulation).

 \item Fig.~5. The average potential energy as a function of temperature.
       The solid, dotted, and dashed curves are obtained from REM2,
       MUCA1, and MUCAREM1, respectively
       (see Tables I and III for the parameters of the simulations).

\end{itemize}
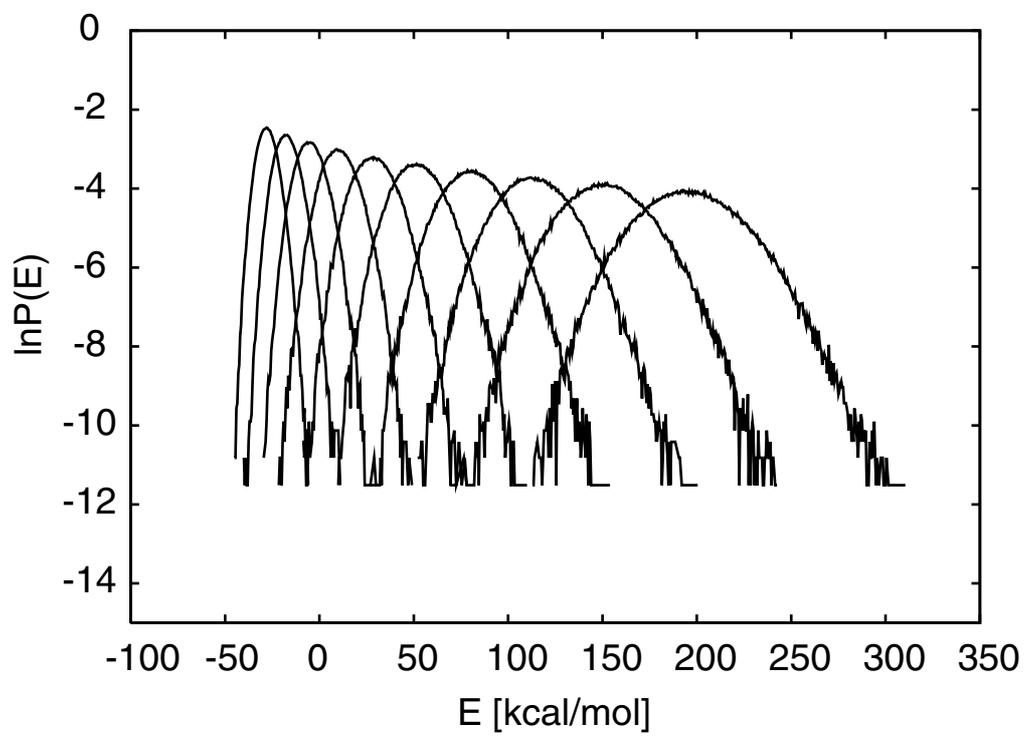
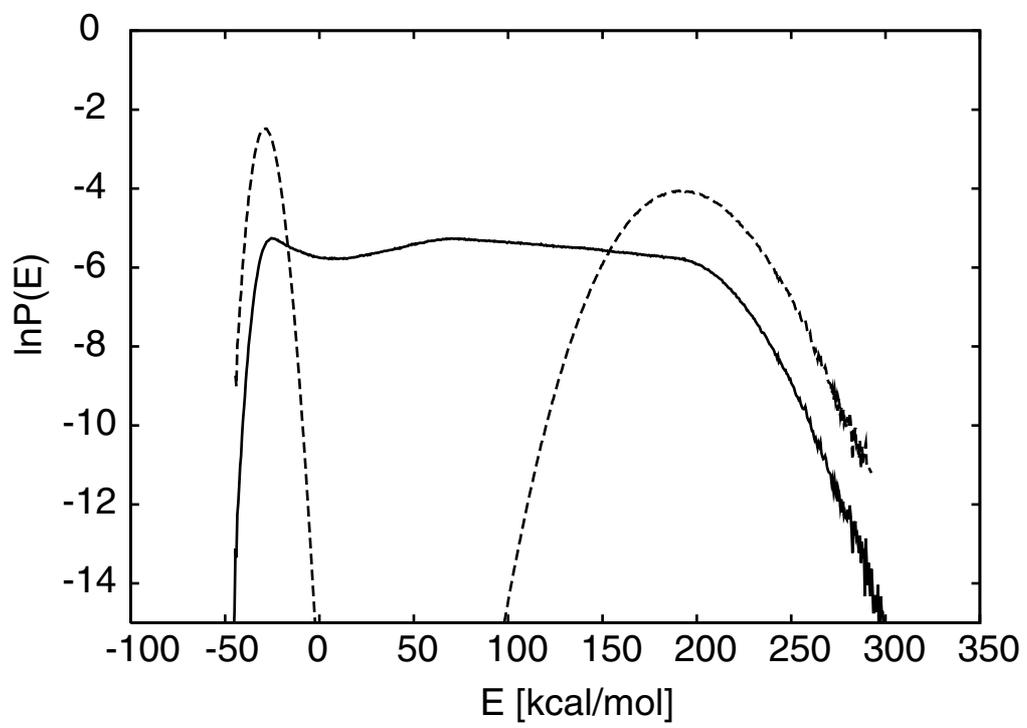
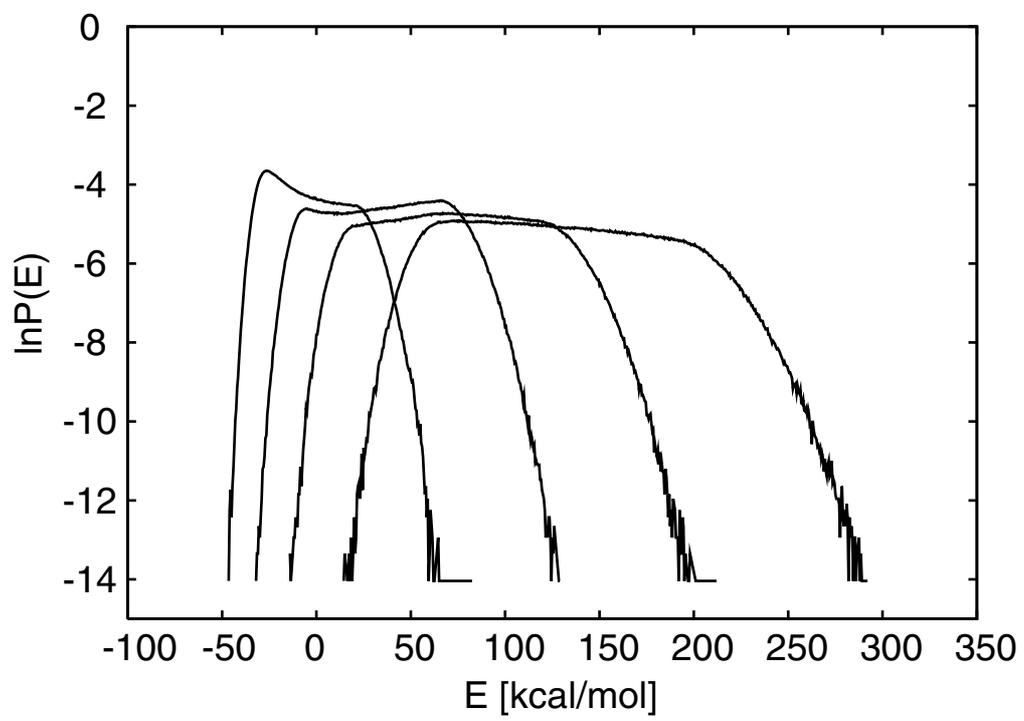
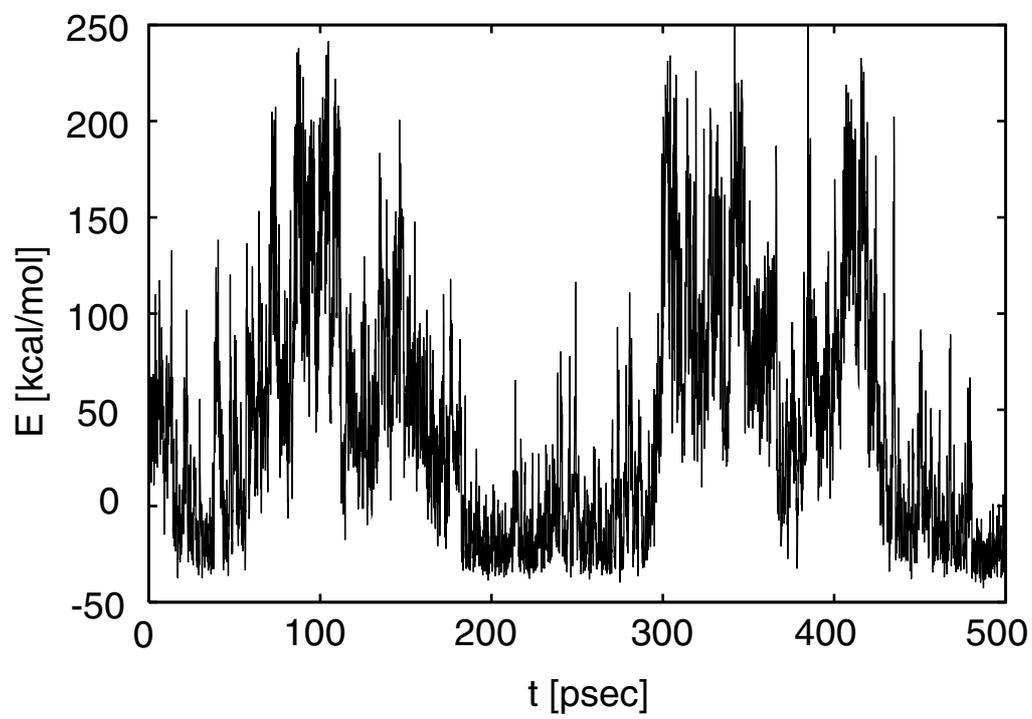
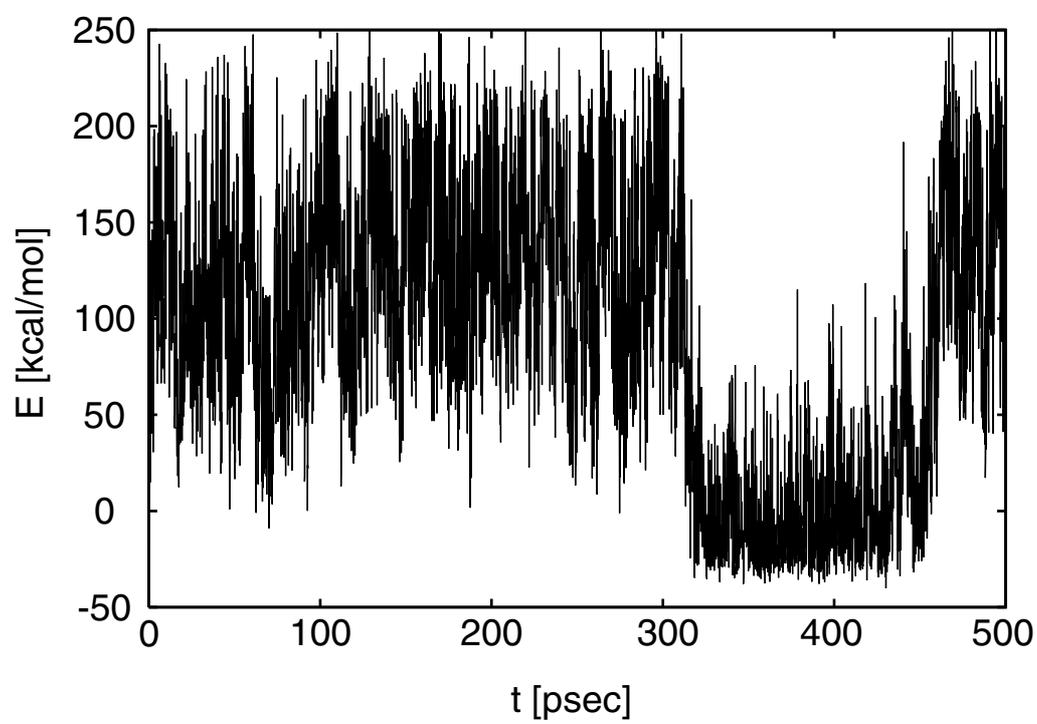
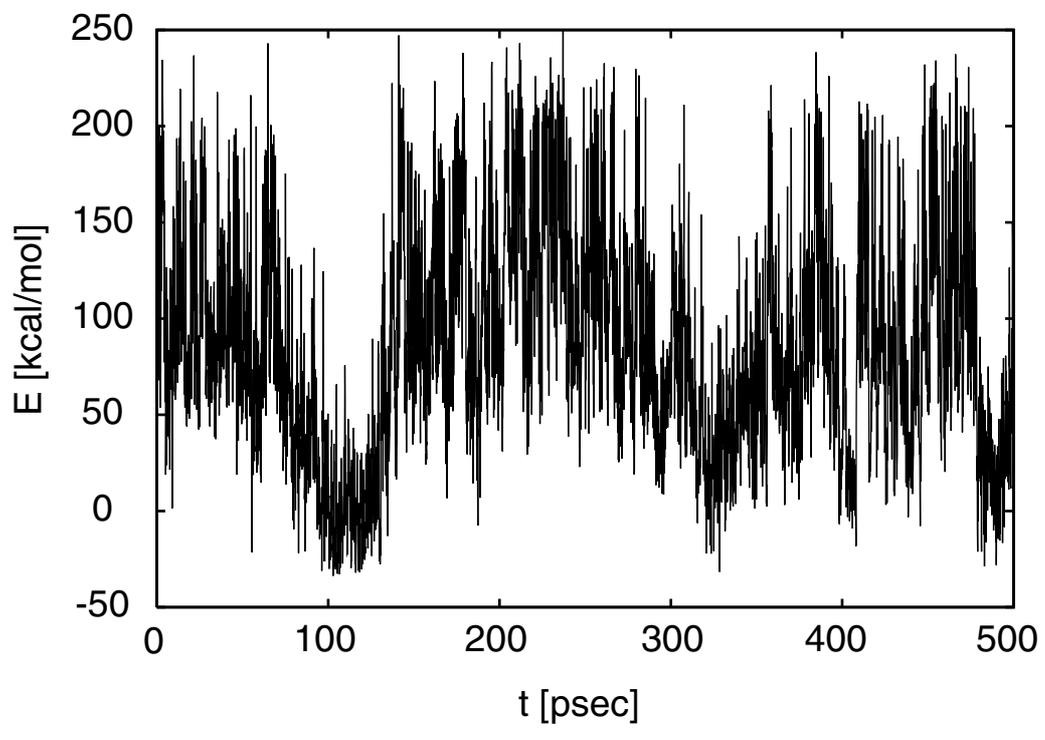
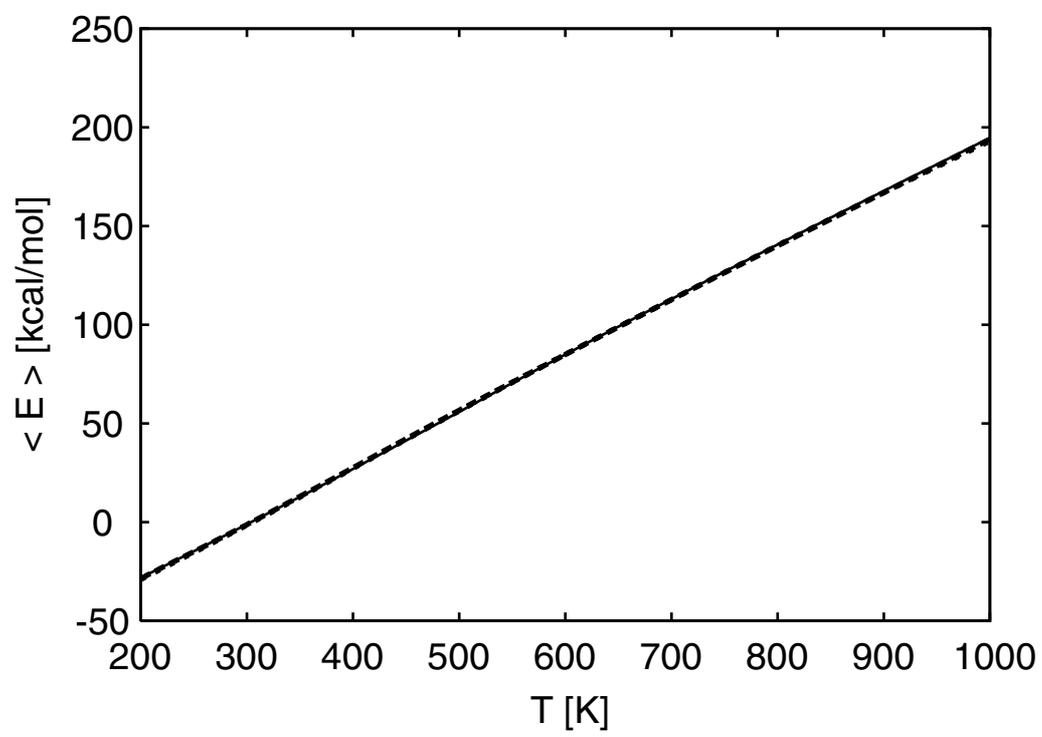
\end{document}